# Line Segment Clipping using Quadrilateral Concavity and Convexity

BIMAL KUMAR RAY
School of Computer Science Engineering & Information Systems
Vellore Institute of Technology, Vellore – 632014, INDIA

**ABSTRACT**
This paper proposes an algorithm for clipping line segment against an axis-aligned rectangular window. The conventional algorithms for line segment clipping treat the clipping boundary and/or the line segment to be clipped as line. The present algorithm treats the clipping boundary and the line segment to be clipped as line segment and using this strategy, it succeeds to avoid computation of false intersection points. A quadrilateral is constructed using the end points of a clipping boundary segment and the end points of the line segment to be clipped as its vertices. The concavity and convexity of the quadrilateral dictates whether a line segment actually intersects the clipping boundary. If the quadrilateral is found to be concave then the line segment is rejected, otherwise the point of intersection of the line segment with the clipping boundary is computed. Since a 'test & intersect' approach is used instead of a 'intersect & test', hence the proposed algorithm does not compute false intersection point thereby reducing the number of divisions required to obtain a clipped line segment. Only one routine can process line segments at any position. Improved performance is observed with respect to the Nicholl-Lee-Nicholl, Liang-Barsky, Cohen-Sutherland and Skala's algorithm through experiments with random line segments using a metric based on execution time.

## I. INTRODUCTION

Removing portions of a line segment that falls outside a window is known as clipping. This paper proposes an algorithm clipping for line segment against a rectangular axis-aligned clipping window. Three classical algorithms for clipping a line segment are Cohen-Sutherland [1], Liang-Barsky [2] and Nicholl-Lee-Nicholl [3].

The Cohen-Sutherland (CS) algorithm computes outcode of the end points of a line segment and based on the outcode it detects line segments that are completely inside the clipping window and removes those that are completely outside the window. The other line segments are tested iteratively and the point of intersection of the line segment with the window boundaries is found until the clipped line segment is found to fall inside the clipping window. The outcode of the end points of the line segment is computed repeatedly as it is intersected against the clipping boundary until the clipped line segment falls inside the window. The algorithm involves iteration for line segments that are neither completely inside nor completely outside the clipping window and for this kind of line segments it computes a maximum of four false intersection points.

The Liang-Barsky (LB) algorithm uses a parametric representation of a line to be clipped in the form
$x = x_1 + (x_2 - x_1)t$ & $y = y_1 + (y - y_1)t$, $0 \leq t \leq 1$.
It computes point of intersection of a line with each of the boundaries of the clipping window. A trivial rejection test is used to reject the lines that do not intersect the window. It is always necessary to find the point of intersection of a line segment (line) with all the clipping boundaries irrespective of whether it actually intersects the window; except for the trivial rejection test. This algorithm too computes a maximum of four false intersection points.

The Nicholl-Lee-Nicholl (NLN) algorithm considers one end point of a line segment and determines the position of the other end point with respect to the clipping window to find out whether it is necessary to compute the point of intersection of the line segment with the clipping boundary and by this process it avoids computing false intersection points. Though the algorithm is highly efficient, but it needs multiple routines to develop, test and validate.

Skala [5] proposed a line and line segment clipping algorithm exploiting position of the vertices of a clipping window with respect to the line or line segment to be clipped. As this paper addresses line segment clipping; the line segment clipping of Skala's approach is briefly reviewed here. The procedure uses CS coding scheme to generate outcode of the end points of the line segment to be clipped apart from a bit vector describing the position of the vertices of the clipping window with respect to the line segment to be clipped. The line segments that are completely inside or completely outside the clipping window are processed using CS outcode whereas the other line segments are clipped using the separation bit vector, a mask called MASK and a table called TAB. The MASK is useful when one of the end points of the line segment is inside the clipping window and the other is outside the clipping window. The table TAB has two vectors TAB1 and TAB2 which is pre-computed and it facilitates detecting the window edge that intersects the line segment. The algorithm can make use of hardware implementation of vector dot and cross product and can exploit parallelism to compute the separation bit vector. The procedure is applicable to convex as well as concave polygonal clipping window and is primarily centered on homogeneous coordinates.

The algorithm proposed in this paper uses concavity and convexity of a quadrilateral to clip a line segment against a rectangular axis-aligned window and unlike the CS and LB algorithm it does not compute false intersection point. Though the NLN algorithm does not compute false intersection point, but it requires a multiple cases to develop, test and validate making it hard to implement. On the contrary, the proposed algorithm needs only two routines to develop. A simple top level routine calls another routine that does the clipping task, at the most twice. In contrast to Skala's algorithm, the proposed procedure uses axis-aligned rectangular clipping window but improved performance is observed with respect to relative execution time.

The CS algorithm treats a line segment to be clipped as a line segment but while computing the point of intersection of a line segment with a clipping boundary it treats the clipping boundary as a line and this is why it computes false intersection points. The LB algorithm too treats a clipping boundary as a line and computes false intersection points. The NLN algorithm treats a clipping boundary as a line segment but it treats a line segment to be clipped as a line. It is observed that it is necessary to treat a line segment as a line segment and the finite clipping boundaries too as finite line segments so that computation of false intersection point(s) of a line segment with the clipping boundaries can be avoided. Skala's algorithm is extended to line segment clipping after it is developed for line clipping.

The proposed algorithm treats a line segment to be clipped as a line segment and so also the clipping boundaries and by this process it succeeds to avoid computation of false intersection points.

## II PROPOSED METHOD – CONCEPT AND IMPLEMENTATION

Consider a rectangular axis-aligned clipping window with its left, right, bottom and top boundary segment defined by the line segment $\{x = x_L, y_B \leq y \leq y_T\}$, $\{x = x_R, y_B \leq y \leq y_T\}$, $\{y = y_B, x_L \leq x \leq x_R\}$ and $\{y = y_T, x_L \leq x \leq x_R\}$ respectively. Let a line segment to be clipped be defined by its end points $A(x_1, y_1)$ and $B(x_2, y_2)$ and the segment $\overline{CD}$ with end points $C(x_L, y_B)$ and $D(x_L, y_T)$ be the left boundary segment of the clipping window. Let us assume that the left end point $A$ of the line segment $\overline{AB}$ lies to the left of the left boundary segment $\overline{CD}$ (Figure 1) and the line segment $\overline{AB}$ is not completely to the left of the left boundary segment. It may be observed from the figure that though the point $B$ is to the right of the bounding segment $\overline{CD}$ but it does not intersect the same. Now join the point $A$ to the point $C$ and $D$ and join the point $B$ to the point $C$ and $D$. This construction reflects treatment of the clipping object as well as the clipping boundary as line segment instead of a line. It may be observed that $ADBCA$ (in this order) forms a concave quadrilateral with angle $C$ as its concave angle.

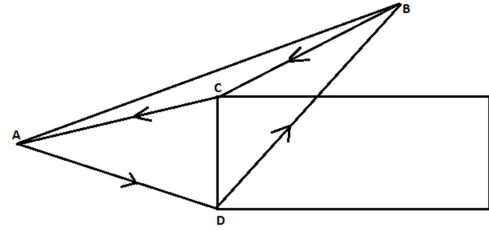

**FIGURE 1** A line segment $\overline{AB}$ that does not intersect the left boundary $\overline{CD}$ of the clipping window

Consider a slightly different arrangement (Figure 2) of the same figure wherein the segment $\overline{AB}$ is below the segment $\overline{CD}$ i. e. both the end points of $\overline{CD}$ are above the line segment $\overline{AB}$. The point $A$ and the point $B$ are joined to the point $C$ and $D$ as before. In this case too,

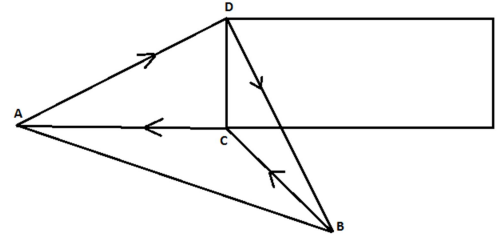

**FIGURE 2** A line segment $\overline{AB}$ below the lower left corner of the clipping window

$ADBCA$ has the shape of a concave quadrilateral with concavity at the vertex $C$.

But if $\overline{AB}$ intersects $\overline{CD}$ (Figure 3) then joining $A$ to the point $C$ and $D$ and $B$ to the point $C$ and $D$, it is observed that $ADBCA$ has the shape of a convex quadrilateral.

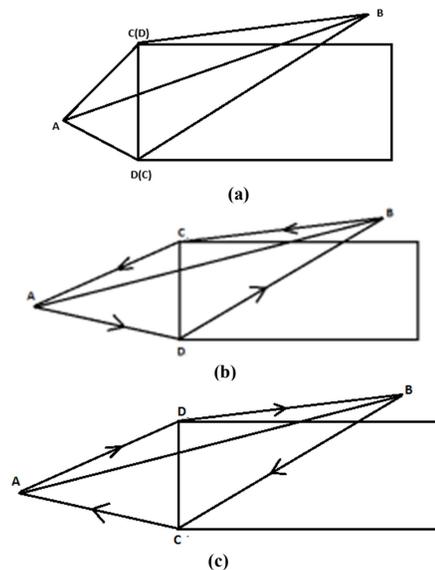

**FIGURE 3** When line segment $\overline{AB}$ intersects $\overline{CD}$ then $ADBCA$ forms a quadrilateral

From the Figures 1 through 3, it is found that if the segment $\overline{AB}$ and $\overline{CD}$ (with the construction defined) form a concave quadrilateral then $\overline{AB}$ cannot intersect $\overline{CD}$ (Figure 1 and 2), but they do intersect when they form a convex quadrilateral (Figure 3) and visa-versa. So in order to determine whether a line segment with one of its end points lying to the left of the left boundary segment and the other point lying to the right of the same boundary segment intersects the left boundary segment then it is necessary and sufficient to test if the $ADBCA$ (in this order) is a convex quadrilateral.

The concavity at $C$ (referring to Figure 1) of the quadrilateral can be computed by determining the sign of the $z$ component of the vector product $\overrightarrow{BC} \times \overrightarrow{CA}$ of the vectors
$\overrightarrow{BC} = (x_L - x_2, y_T - y_2)$ and $\overrightarrow{CA} = (x_1 - x_L, y_1 - y_T)$
following the right hand thumb rule. It is evident from the Figure 1 that the $z$ component of the vector product $\overrightarrow{BC} \times \overrightarrow{CA}$ defined by
$z_T = (x_L - x_2)(y_1 - y_T) - (y_T - y_2)(x_1 - x_L)$ is negative because the vector cross product $\overrightarrow{BC} \times \overrightarrow{CA}$ forms a left handed system as seen from the front. Thus the quadrilateral is concave at the vertex $C$ if $z_T < 0$ and the line segment $\overline{AB}$ is rejected. Similarly, referring to Figure 2,
$\overrightarrow{BC} = (x_L - x_2, y_B - y_2)$ and $\overrightarrow{CA} = (x_1 - x_L, y_1 - y_B)$
and the vector product $\overrightarrow{BC} \times \overrightarrow{CA}$ forms a right handed system as seen from the front and hence the $z$ component of the vector product $\overrightarrow{BC} \times \overrightarrow{CA}$ defined by
$z_B = (x_L - x_2)(y_1 - y_B) - (y_B - y_2)(x_1 - x_L)$
is positive. Thus the line segment $\overline{AB}$ is rejected when $z_B > 0$. The line segment $\overline{AB}$ intersects the bounding segment $\overline{CD}$ as shown in Figure 3 when neither of the expressions $z_T < 0$ and $z_B > 0$ turns out to be true. The figure 3(a) shows a convex quadrilateral as the line segment $\overline{AB}$ intersects the bounding segment $\overline{CD}$. The Figure 3(b) shows how the angle at the top-left corner is detected as convex and the Figure 3(c) shows the same at the bottom-left corner. The Figure 3(b) shows that the vector cross product $\overrightarrow{BC} \times \overrightarrow{CA}$ forms a right handed system and the Figure 3(c) shows that it forms a left handed system. Thus whenever $z_T < 0$ and $z_B > 0$ turn out to be false, it is necessary to determine the point of intersection of the segment $\overline{AB}$ with the segment $\overline{CD}$ and the point of intersection is defined by
$\left(x_L, y + \frac{(y_2 - y_1)(x_1 - x_L)}{x_2 - x_1}\right)$.
Though the factor $x_2 - x_1$ appears in the denominator; it does not cause arithmetic overflow because if the line segment to be clipped is vertical then both its end points will be to the left of the left boundary segment and will be rejected by carrying out an additional test $x_2 < x_L$ after $x_1 < x_L$ succeeds.

The Figure 4 shows the case where the end point $A$ of the line segment $\overline{AB}$ to be clipped falls to the right of the right boundary $\overline{CD}$ (using the same symbol for right boundary segment as for the left boundary segment for ease of comprehension) of the clipping window and is above the top-right corner of the clipping window but the line segment is not trivially rejected. The boundary segment $\overline{CD}$ is defined by the points $C(x_R, y_B)$ and $D(x_R, y_T)$. The vector product $\overrightarrow{BC} \times \overrightarrow{CA}$ of the vectors $\overrightarrow{BC} = (x_R - x_2, y_T - y_2)$ and $\overrightarrow{CA} = (x_1 - x_R, y_1 - y_T)$ forms a right handed system as seen from the front and hence the $z$ component of the vector product $\overrightarrow{BC} \times \overrightarrow{CA}$ defined by
$z_T = (x_R - x_2)(y_1 - y_T) - (y_T - y_2)(x_1 - x_R)$
is positive.

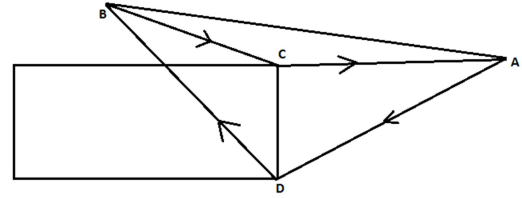

**FIGURE 4** The segment $\overline{AB}$ does not intersect the right boundary $\overline{CD}$ and is above the top-right corner of the window

The Figure 5 shows the case where the end point $A$ of the line segment $\overline{AB}$ to be clipped falls to the right of the right boundary segment $\overline{CD}$ and is below the bottom-right corner of the clipping window but the line segment is not trivially rejected. The boundary segment $\overline{CD}$ is defined by the points $C(x_R, y_B)$ and $D(x_R, y_T)$. The vector product $\overrightarrow{BC} \times \overrightarrow{CA}$ of the vectors $\overrightarrow{BC} = (x_R - x_2, y_B - y_2)$ and $\overrightarrow{CA} = (x_1 - x_R, y_1 - y_B)$ forms a left handed system as seen from the front and hence the $z$ component of the vector product $\overrightarrow{BC} \times \overrightarrow{CA}$ defined by
$z_B = (x_R - x_2)(y_1 - y_B) - (y_B - y_2)(x_1 - x_R)$
is negative.

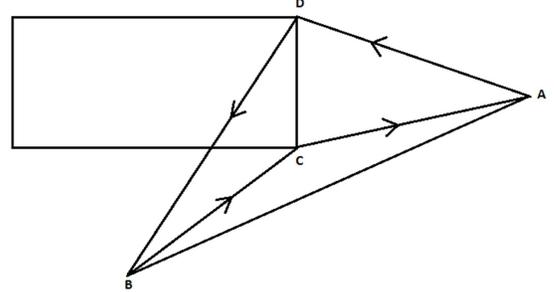

**FIGURE 5** The segment $\overline{AB}$ does not intersect the right boundary $\overline{CD}$ and is below the bottom-right corner of the window

The Figure 6 shows the case where the line segment $\overline{AB}$ intersects the right boundary. The quadrilateral $ADBCA$ is convex now. The figure 6(a) shows the convex quadrilateral, 6(b) shows how to detect the convexity at the top-right corner and 6(c) shows the same for the bottom-right corner.

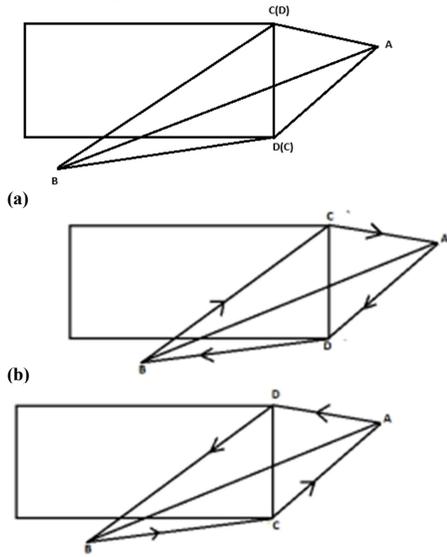

(a)

(b)

(c)

**FIGURE 6** When line segment $AB$ intersects $CD$ then $ADBCA$ forms a quadrilateral

The cases where the end point $A$ is either below the bottom boundary or above the top boundary and the line segment is not rejected trivially, can similarly be addressed. The necessary pictorial representations of these cases are shown in Figure 7 and 8.

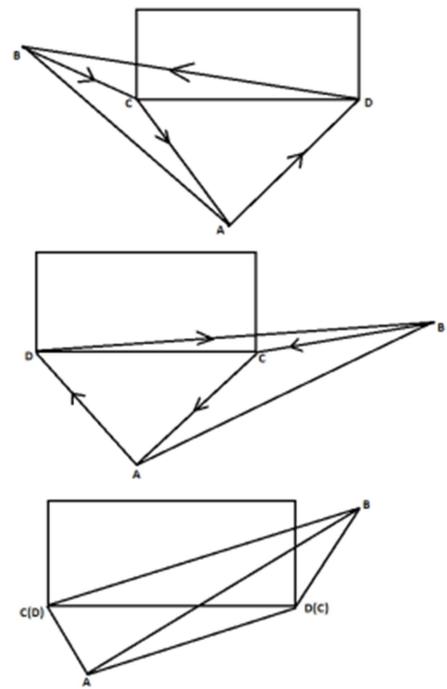

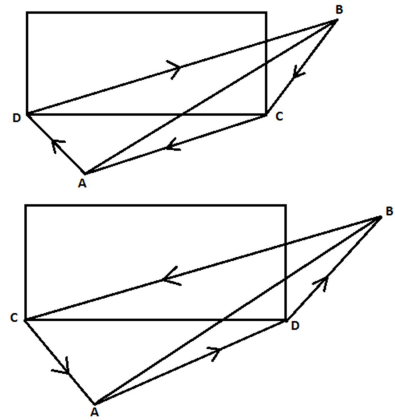

**FIGURE 7** The end point $A$ is below the bottom boundary but the line segment $\overline{AB}$ is not trivially rejected

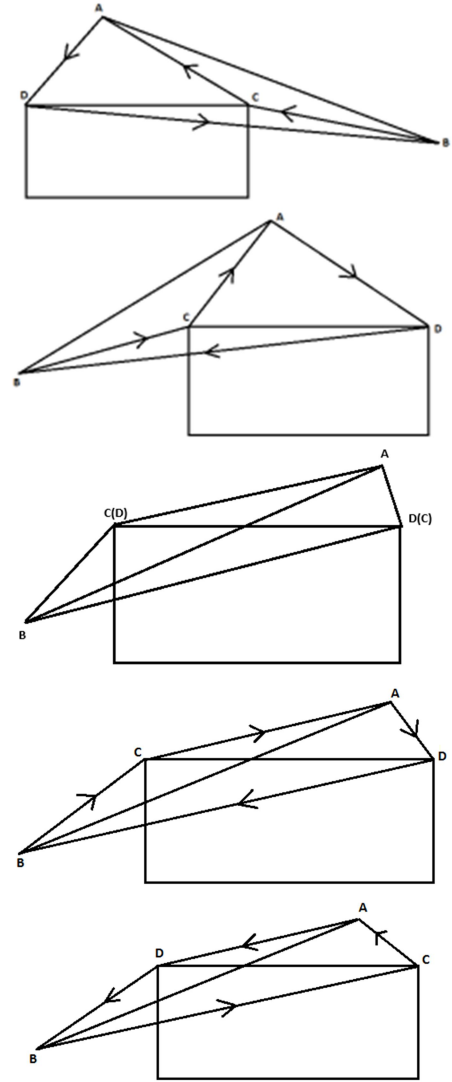

**FIGURE 8** The end point $A$ is above the top boundary but the line segment $\overline{AB}$ is not trivially rejected

A line segment with both end points lying inside the clipping window survives the clipping test and so it is eventually accepted. The following table (Table I) shows the position of the end point $A$ with respect to the window boundary and the predicate that leads to rejection of the line segment.

TABLE I Predicate used to facilitate clipping for different non-trivial position of the line segment

| Position of the end point $A(x_1, y_1)$ | Rejection predicate |
|---|---|
| $x_1 < x_L$ and the line segment is above the top-left corner | $(x_L - x_2)(y_1 - y_T) - (x_1 - x_L)(y_T - y_2) < 0$ |
| $x_1 < x_L$ and the line segment is below the bottom-left corner | $(x_L - x_2)(y_1 - y_B) - (x_1 - x_L)(y_B - y_2) > 0$ |
| $x_1 > x_R$ and the line segment is above the top-right corner | $(x_R - x_2)(y_1 - y_T) - (x_1 - x_R)(y_T - y_2) > 0$ |
| $x_1 > x_R$ and the line segment is below the bottom-right corner | $(x_R - x_2)(y_1 - y_B) - (x_1 - x_R)(y_B - y_2) < 0$ |
| $y_1 < y_B$ and the line segment is below the bottom-right corner | $(x_L - x_2)(y_1 - y_B) - (x_1 - x_L)(y_B - y_2) < 0$ |
| $y_1 < y_B$ and the line segment is below the bottom-left corner | $(x_R - x_2)(y_1 - y_B) - (x_1 - x_R)(y_B - y_2) > 0$ |
| $y_1 > y_T$ and the line segment is above the top-right corner | $(x_L - x_2)(y_1 - y_T) - (x_1 - x_L)(y_T - y_2) > 0$ |
| $y_1 > y_T$ and the line segment is below the top-left corner | $(x_R - x_2)(y_1 - y_T) - (x_1 - x_R)(y_T - y_2) < 0$ |

The method is output directed. It does not compute false intersection point. The number of intersection points computed is exactly the same as the number of intersection points of the output line segment. The CS algorithm, on the other hand, computes a maximum of four false intersections and so also the LB algorithm. Though the NLN algorithm does not compute false intersection, but it needs multiple routines to develop, test and validate. Skala's algorithm too does not compute false intersection and avoids division using homogeneous coordinates system but the proposed algorithm is observed to exhibit better performance in terms of execution speed.

The procedure called **Clip** presented below is used to clip a line segment against a rectangular window with respect to one end point of the line segment. It takes the coordinates $(x_1, y_1)$ and $(x_2, y_2)$ of the end points of the line segment to be clipped and the window parameters $x_L$, $x_R$, $y_B$ and $y_T$ as input parameters and treats one of the end points $(x_1, y_1)$ of the line segment as output parameter as well. A variable *intersect* stores display information namely, 1 for display and 0 for no display.

```
Procedure Clip(x₁, y₁, x₂, y₂, xL, xR, yB, yT, intersect)
begin
  if x₁ < xL then
    begin
      if x₂ < xL then return 0;
      if (xL − x₂)(y₁ − yT) < (x₁ − xL)(yT − y₂) then
        intersect := 0;
      else if (xL − x₂)(y₁ − yB) > (x₁ − xL)(yB − y₂) then
        intersect := 0;
      else
        begin
          y₁ := y₁ + (y₂ − y₁)(xL − x₁)/(x₂ − x₁);
          x₁ := xL;
          intersect := 1;
        end;
    end;
  else if x₁ > xR then
    begin
      if x₂ > xR then return 0;
      if (xR − x₂)(y₁ − yT) > (x₁ − xR)(yT − y₂)
        then intersect := 0;
      else if (xR − x₂)(y₁ − yB) < (x₁ − xR)(yB − y₂)
        then intersect := 0;
      else
        begin
          y₁ := y₁ + (y₂ − y₁)(xR − x₁)/(x₂ − x₁);
          x₁ := xR;
          intersect := 1;
        end;
    end;
  else intersect := 1;
  if y₁ < yB then
    begin
      if y₂ < yB then return 0;
      if (xL − x₂)(y₁ − yB) < (x₁ − xL)(yB − y₂) then
        intersect := 0;
      else if (xR − x₂)(y₁ − yB) > (x₁ − xR)(yB − y₂) then
        intersect := 0;
      else
        begin
          x₁ := x₁ + (x₂ − x₁)(yB − y₁)/(y₂ − y₁);
          y₁ := yB;
          intersect := 1;
        end;
    end;
  else if y₁ > yT then
    begin
      if y₂ > yT then return 0;
      else if (xL − x₂)(y₁ − yT) > (x₁ − xL)(yT − y₂) then
        intersect := 0;
      else if (xR − x₂)(y₁ − yT) < (x₁ − xR) (yT − y₂) then
        intersect := 0;
      else
        begin
          x₁ := x₁ + (x₂ − x₁)(yB − y₁)/(y₂ − y₁);
          y₁ := yT;
          intersect := 1;
        end;
    end;
  else
    intersect := 1;
  return intersect;
end.
```

A higher level routine called ***ClipSegment***, displayed below accepts the coordinates of the line segment as well as the window parameters as its arguments wherein the end points of the line segment act as input as well as output parameter.

Procedure ***ClipSegment*** ($x_1, y_1, x_2, y_2, x_L, x_R, y_B, y_T$)
begin
 if Clip($x_1, y_1, x_2, y_2, x_L, x_R, y_B, y_T$, intersect) = 0 then exit;
 if Clip($x_2, y_2, x_1, y_1, x_L, x_R, y_B, y_T$, intersect) = 0 then exit;
 else display the clipped line segment;
end;

If both the end points of the line segment are outside the clipping window then the first call to the routine **Clip** inside the **ClipSegment** routine rejects the line segment. If a point $A(x_1, y_1)$ is inside the clipping window then the predicates $x_1 < x_L$, $x_1 > x_R$, $y_1 < y_B$ and $y_1 > y_T$ turn out to be false and the display information is stored as 1 in the intersect variable of the procedure **Clip**. The second call to the same procedure **Clip** inside the procedure **ClipSegment** treats the other end point $B(x_2, y_2)$ of the line segment in a similar manner; if it is inside the clipping window and the line segment is displayed.

Some singular cases are shown in Figure 9 wherein both end points of the line segment fall on the boundary of the clipping window (top left), the line segment is completely inside the clipping window (top right), the line segment is on the upper edge of the window (bottom left) and the line segment is on the right edge of the window.

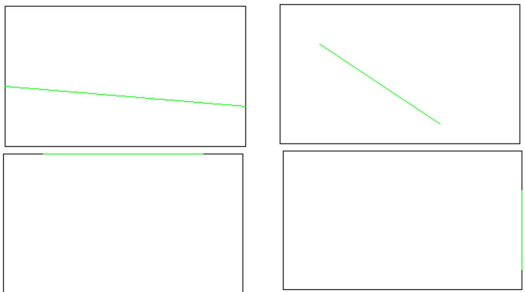

**FIGURE 9** End points of a line segment falling on the clipping boundary (top left), both end points falling inside the clipping window (top right right), the line segment is on the upper edge of the clipping window (bottom left) and the line segment is on a side edge of the clipping window (bottom right)

The Figure 10 shows experimental output of clipping ten (top figure), one hundred (middle figure) and one thousand (bottom figure) random line segments. The input line segments are shown in blue and the clipped line segments are in green. The clipping window is drawn with black boundary.

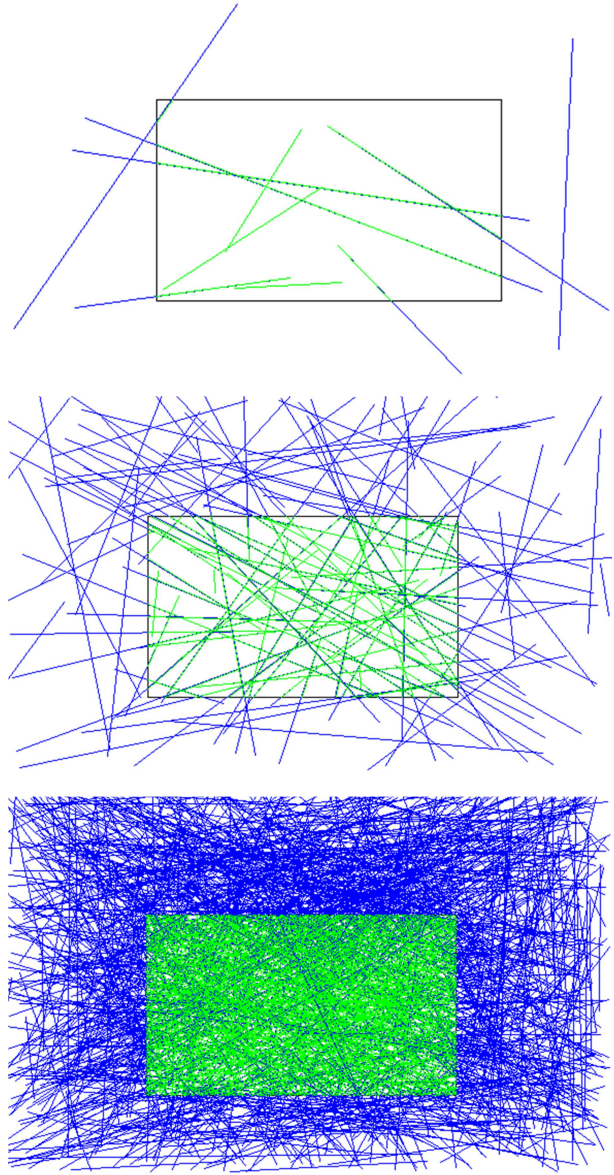

**FIGURE 10** Experiment with ten (top), one hundred (middle) and one thousand (bottom) random line segments

### III COMPARISONS
The proposed algorithm is tested with respect to the NLN, LB, CS and Skala's algorithm using one hundred iterations of a number of random line segments varying from ten to ten million. The implementation in [1] is used for the CS and the LB algorithm and that in [3] is used for the NLN algorithm. The Skala's algorithm is coded from [5]. The same set of random line segments is used for the proposed as well as the benchmark algorithms in one pass of the iteration. But a different set of random line segments is used in different pass of the iteration. The average of the total execution time of the NLN, LB, CS and Skala's algorithm is compared with the average of the total execution time of the proposed algorithm using a metric defined by

*Average Relative Exection time =*
$\frac{Average\ of\ Total\ execution\ time\ using\ a\ benchmark\ algorithm}{Average\ of\ Total\ execution\ time\ of\ the\ proposed\ algorithm}$.

The higher is the value of the metric the better is the performance of the proposed algorithm with respect to the NLN, LB, CS and Skala's algorithm. The algorithms are implemented in C++ and are executed on an HP system with 11th Generation Intel Core i5-1135G7@2.40 GHz processor and 8GB RAM; measuring execution time in milliseconds and the results are shown in Table II.

**TABLE II** Average of relative total execution time

| No. of LS | NLN/ClipSg | LB/ClipSg | CS/ClipSg | Skala/ClipSg |
|---|---|---|---|---|
| 10 | 1.2532 | 1.3665 | 1.2919 | 1.6766 |
| 100 | 1.2154 | 1.2745 | 1.1884 | 1.5940 |
| 1000 | 1.2227 | 1.4713 | 1.2266 | 1.6072 |
| 10000 | 1.1571 | 1.4241 | 1.1833 | 1.5516 |
| 100000 | 1.1684 | 1.4357 | 1.1860 | 1.5492 |
| 1000000 | 1.1653 | 1.4472 | 1.1945 | 1.5617 |
| 10000000 | 1.1651 | 1.4452 | 1.1941 | 1.5624 |
| **Average** | **1.1924** | **1.4092** | **1.2092** | **1.5861** |

The first column of the table shows the number of random line segments and the other four columns show the average relative total execution time over one hundred iterations as produced by the benchmark algorithms namely, NLN (second column; named **NLN/ClipSg**), LB (third column; named **LB/ClipSg**), CS (fourth column; named **CS/ClipSg**) and Skala (fifth column; named **Skala/ClipSg**) to the proposed algorithm. The coordinates used for NLN, LB and CS are mapped to homogeneous coordinates for testing Skala's algorithm. The results in the table show that the proposed algorithm exhibits better performance than the NLN, LB, CS and Skala's algorithm. The last row of the table shows the overall average relative performance of the proposed algorithm.

The maximum number of false intersection points computed by the CS and LB algorithm is four but the present algorithm, like the NLN, does not compute false intersection point. The number of divisions performed is exactly the same as the number of intersection points.

## IV CONCLUSION

This paper uses concavity and convexity of a quadrilateral constructed with the clipping line segment and clipping window boundary segment to design an algorithm for line segment clipping against an axis-aligned rectangular window and develop and test it. It shows performance improvement over the NLN, LB, CS and Skala's algorithm through experiments in multiple iterations using sets of random line segments; each set containing a different number of random line segments. The algorithm is simpler than the NLN algorithm in that development, testing and validation of the proposed algorithm is easier than the NLN algorithm and unlike the LB and CS algorithm; it does not compute false intersection points.

## REFERENCES

[1] J. D. Foley, A. Van Dam, S. K. Feigner, and J. F. Hughes. Computer Graphics: Principles and Practice. Addison-Wesley (2nd edition in C), 1996.
[2] Y. D. Liang, and B. A. Barsky. "A New Concept and Method for Line Clipping". ACM Transactions on Graphics 3 : 1 (1984), pp. 1 – 22.
[3] T. M. Nicholl, D. T. Lee, and R. A. Nicholl. "An Efficient New Algorithm for 2-D Line Clipping". Computers & Graphics 21: 4(1987), pp. 253 – 262.
[4] David, F. Rogers. Procedural Elements for Computer Graphics, 2nd edition Tata McGraw-Hill, 2005.
[5] Vaclav Skala, "A new approach to line and line segment clipping in homogeneous coordinates", The Visual Computer, 21 (2005), pp. 905 – 914